# Distinctive Signature of Indium Gallium Nitride Quantum Dot Lasing in Microdisks Cavities

Alexander Woolf[a], Tim Puchtler[b], Igor Aharonovich[c], Tongtong Zhu[b], Nan Niu[a], Danqing Wang[a], Rachel A. Oliver[b], and Evelyn L. Hu[a]

[a]School of Engineering and Applied Sciences, Harvard University, Cambridge, Massachusetts 02138, USA; [b]Department of Material Science and Metallurgy, University of Cambridge, 27 Charles Babbage Road, Cambridge CB3 0FS, United Kingdom; [c]School of Physics and Advanced Materials, University of Technology Sydney, Ultimo 2007, New South Wales, 2007 Australia

**Abstract**
Low threshold lasers realized within compact, high quality optical cavities enable a variety of nanophotonics applications. Gallium nitride (GaN) materials containing indium gallium nitride (InGaN) quantum dots and quantum wells offer an outstanding platform to study light matter interactions and realize practical devices such as efficient light emitting diodes and nanolasers. Despite progress in the growth and characterization of InGaN quantum dots, their advantages as the gain medium in low threshold lasers have not been clearly demonstrated. This work seeks to better understand the reasons for these limitations by focusing on the simpler, limited-mode microdisk cavities, and by carrying out comparisons of lasing dynamics in those cavities using varying gain media including InGaN quantum wells, fragmented quantum wells, and a combination of fragmented quantum wells with quantum dots. For each gain medium, we utilize the distinctive, high quality (Q~5500) modes of the cavities, and the change in the highest-intensity mode as a function of pump power to better understand the dominant radiative processes. The variations of threshold power and lasing wavelength as a function of gain medium help us identify the possible limitations to lower-threshold lasing with quantum dot active medium. In addition, we have identified a distinctive lasing signature for quantum dot materials, which consistently lase at wavelengths shorter than the peak of the room temperature gain emission. These findings not only provide better understanding of lasing in nitride-based quantum dot cavity systems, but also shed insight into the more fundamental issues of light-matter coupling in such systems.

**Significance statement**
The III-nitride family of materials has already demonstrated tremendous optical efficiency and versatility for devices across a broad range of wavelengths.
Quantum dots formed in these materials, with advantages such as improved carrier confinement should offer even greater device efficiency. They are also important constituents for fundamental studies of light-matter interaction. However, that promise has been far from realized, and this is a complex problem to address. This work, through a comparative study of quantum dot, quantum well and fragmented quantum well gain media in compact microdisk cavities, allows better understanding of the limitations to lasing for the quantum dot samples. These results allow both improved device efficiency as well as fundamental insights into quantum dot-cavity interactions in these materials.

**Main text**

The family of III-nitrides materials is promising for the realization of photonic devices including light emitting diodes and lasers (1-6). These systems have also been explored to engineer quantum emitters based on nitride quantum dots (QDs) that are active at room temperature and are suitable for applications in quantum information science (7). The increased control of the growth of isolated InGaN QDs in a high quality GaN matrix has also resulted in a fabrication of high quality nitride microcavities including microdisks and photonic crystal cavities with emission at wavelengths in the blue spectral range (320nm-470nm), (8,9). Subsequently, electrically excited emitters and low threshold lasers based on QDs as the gain medium have been studied (10, 11). However, in spite of the theoretical advantages of QD lasers, related to the density of electronic states for QDs (12, 13, 14, 15), QD-microcavity devices still have higher thresholds than QW-microcavity devices for the nitride materials (8, 9, 16). Furthermore, InGaN QDs formed through a modified droplet epitaxy (MDE) method are always associated with an accompanying fragmented quantum well (fQW) layer. It is thus important to determine the possible influence of the fQW layer on lasing properties in order to achieve a better understanding of the unique contribution from the QDs to the lasing mechanism.

This work studies the lasing dynamics and the correlation between lasing threshold and lasing wavelength of microdisk lasers whose active areas comprise either QWs, fQWs or a combination of QDs with fQWs. Through a detailed comparison of lasing in devices with the same geometries but with different active areas, we find that the distinctive signature of QD facilitated gain is lasing at shorter wavelengths than the average of the background emission. The short wavelength emission is further confirmed by low-temperature, low-power photoluminescence (PL) measurements.

**Results**

**Device fabrication**

To investigate the dynamics of InGaN QD microcavity lasers we simultaneously fabricated numerous 1.2 μm diameter microdisk cavities from four wafers with an identical cavity design except for the three active layers embedded in *c*-plane GaN. The active layers were either composed of $In_xGa_{1-x}N$ QWs, $In_xGa_{1-x}N$ fQWs, or $In_xGa_{1-x}N$ QDs + $In_xGa_{1-x}N$ fQWs, respectively. Furthermore, the two QD+fQW samples (A and B) differed in the average density of QDs, and in the percentage of fQW comprising the active layer. The volume of InGaN QW in each sample was analyzed by an atomic force microscope (AFM), with the percentage of the surface covered by the QW layer as: 100% (QW), 78.3% (fQW), 75.3% (sample A), and 63.9% (Sample B). The QD density is estimated to be $2.8\pm0.1 \cdot 10^{10}$ cm$^{-2}$ and $2.5\pm0.1 \cdot 10^{9}$ cm$^{-2}$ for samples A and B respectively. To ensure that all samples exhibit similar room temperature PL peak emission wavelengths (450 ± 5 nm), the InGaN growth temperature was varied between 707 to 755 °C to compensate for any loss of indium during annealing steps. Further details related to the material growth and characterization are described in *Materials and Methods*.

Microdisk cavities with equal diameters were consequently fabricated from these materials using reactive ion etching, and photoelectrochemical etching to achieve the final undercut structure of the microdisk. The fabrication of the microdisk cavities was identical to that described in previous work (2). The microdisks were designed to be 1.2 μm in diameter and 200 nm in thickness in order to have multiple modes widely spaced over the entire emission spectrum. Figures 1 (a) and (b) show representative SEM images of the completed microdisk cavities. Figures 1 (c), (d), and (e) are schematics of the active layers in the QW, fQW, and fQW+QD devices respectively.

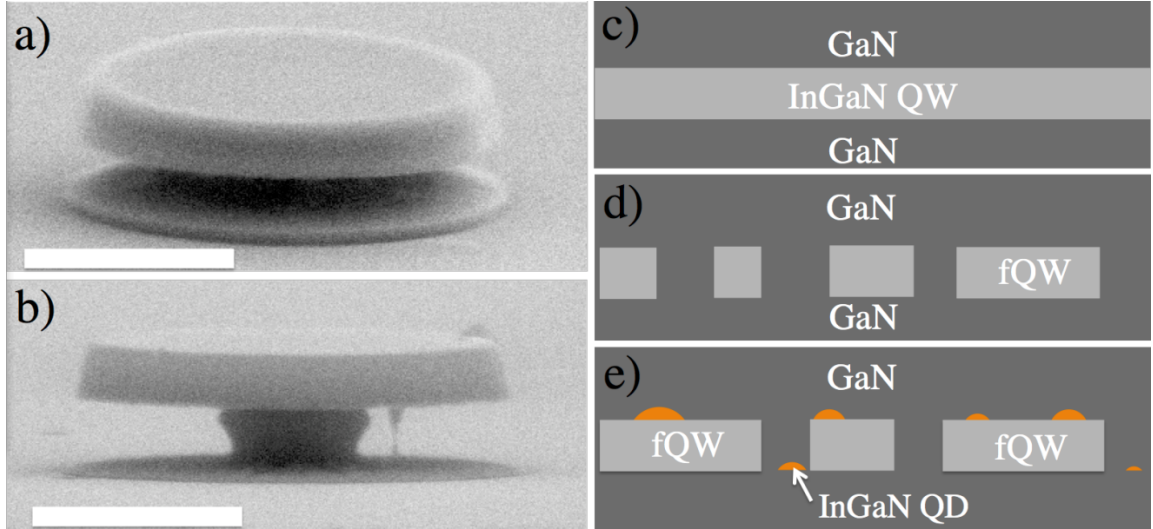

Fig 1: Images and schematic of devices examined in this work. a,b) SEM image of the top and side view of a completed 1.2μm GaN microdisk cavity. Scale bar is 500nm in width. c, d, e) Schematic of one of the three active layers embedded in the QW, fQW and fQW+QD samples respectively.

**Optical measurements**

Room temperature PL measurements were carried out to measure the quality factor of the microdisk cavities. Quality factors as high as 5500 were measured from all samples, which is comparable with the best nitride microdisks currently achievable at these cavity dimensions (17). The devices were measured using a 100x objective in a standard micro-photoluminescence (u-PL) setup in which a frequency-doubled pulsed titanium sapphire laser emitting at 380nm was used in order to isolate the InGaN layers (bandgap ~425nm) from the GaN bulk emission (bandgap 360nm).

Low temperature (4K), low power (1.4 μW) excitation measurements of fQW+QD microdisks suggest that the QDs predominantly emit light on the short wavelength side of the gain spectrum. These data are shown in Figure 2b, as well as data taken at a higher pump power (43 μW). Of all the active layer material considered here, only the fQW+QD A and B samples exhibited sharp peaks, identified as InGaN QD excitons in previous work (18, 19, 20), which were consistently located on the short wavelength side of the emission spectrum and visible up to 40K. The short wavelength QD emission is believed to be a consequence of the MDE technique, as has been previously explained (21). The excitons were observable only at the lowest incident pump powers, while the modes were clearly seen at both high and low pump powers. The dominant modes do not appreciably change as the pump power is raised from 1.4 to 43 μW power, but the gain spectrum does appear to shift by approximately 10 nm to shorter wavelengths, as can be seen in the Gaussian fits to the data.

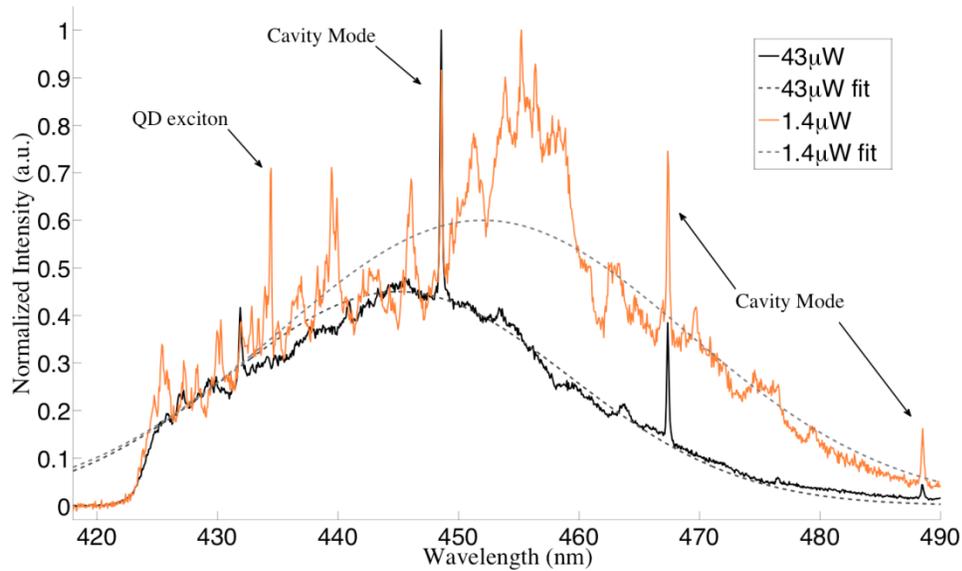

*Fig 2: Normalized PL spectra from a fQW+QD (A) microdisk obtained at 4K. The exciton peaks are only visible at low pump powers (up to ~ 10uW) whereas the cavity modes are visible at all pump powers. Gaussian fits (dashed lines) show that the emission spectrum shifts to shorter wavelengths with increasing pump power.*

    Light in-Light Out (L-L) lasing curves were obtained at room temperature from 20 microdisks on each of the four wafers, which allowed for the identification of the laser threshold as well as lasing wavelength. Representative room temperature microdisk lasing spectra are shown in Figure 3 (a), (b) and (c) for a QW, fQW, and QD+fQW device respectively, with full L-L curves displayed in the insets. The log-log plot clearly indicates the evolution from spontaneous emission to lasing.

    Particularly notable in the data of Figure 3 is the transformation of the PL spectra for the three samples under progressively higher pump powers. Although there were across-sample variations in the widths and peak positions of the luminescence, in general at low input powers, all spectra display a broad luminescence peak, characteristic of the inhomogeneous widths of the QW and fQWs. Against these broad backgrounds are the distinctive, narrow peaks that modulate the broad PL spectra. These correspond to the modes of the microdisks, indicative of the coupling of the emitters to the surrounding cavity. The most prominent modes help to spotlight the optical transitions that are best coupled to the cavity at a given incident pump power, whether because of advantages in radiative emission rate, spatial overlap with a mode or resonance in frequency.

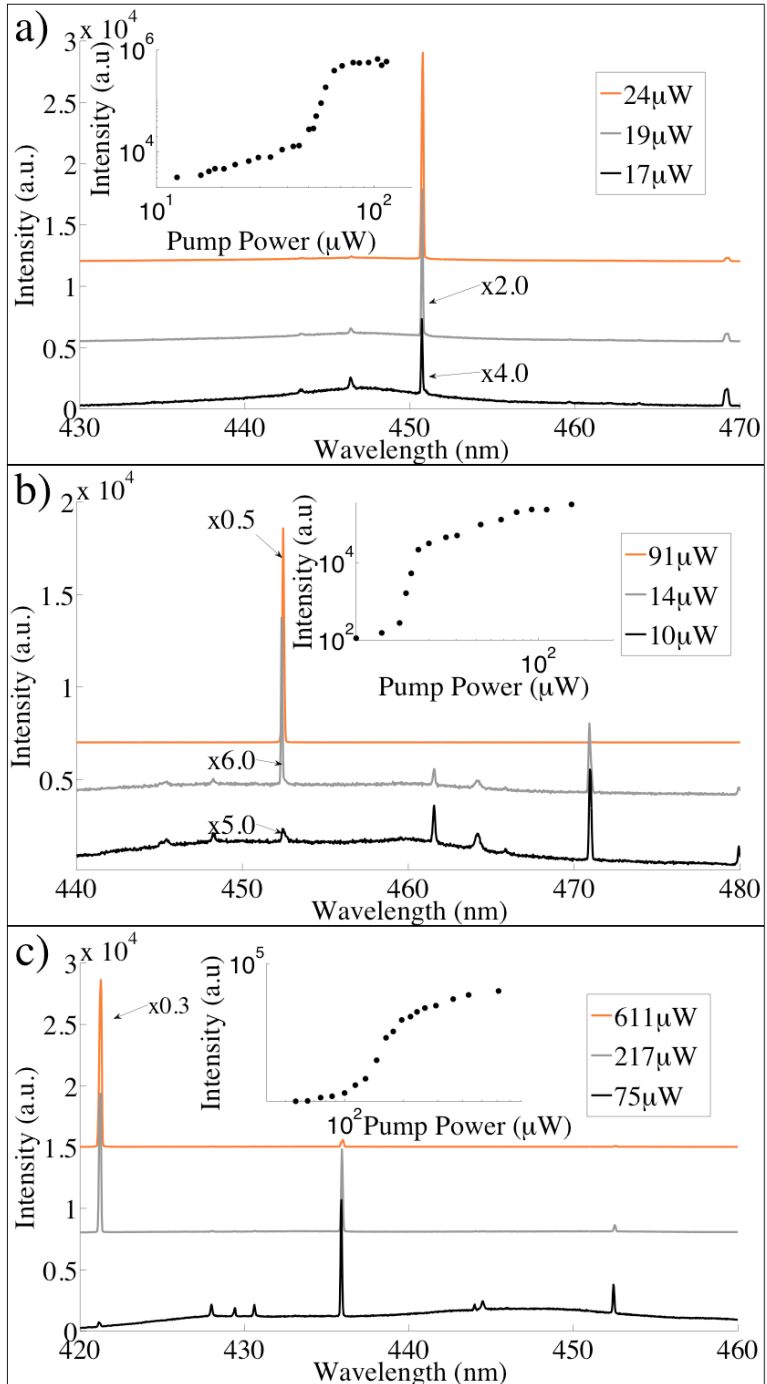

*Fig 3: Lasing spectra from devices with varying gain media. a, b, c) Typical room temperature lasing spectra of QW, fQW, and QD+fQW samples respectively. Inset: full log-log lasing curves showing the three laser regimes of spontaneous emission, amplified spontaneous emission, and lasing.*

For the QW the most prominent mode appears near the peak of the PL distribution; this mode continues its prominence under higher pump powers, and ultimately defines the lasing wavelength. For the fQW sample, the dominant modes under low pump power are those at the long-wavelength tail of the distribution (~ 471 nm). At a higher pump power, the dominant mode is blue-shifted (~ 452 nm), closer to the center of the broad PL peak. Similar to the QW samples, lasing takes place at a wavelength close to the center of the broad gain spectrum. The PL spectra of the QD+fQW samples initially have similarities to the fQW (alone) samples: at the lowest pump power shown (75 µW), the spectrum in Figure 4(c) displays a prominent mode at the approximate center of the gain distribution (~ 436 nm), with a small red-shifted mode at ~ 452 nm. The distinctive change for the QD+fQW samples, at the higher pump powers, is the dramatic dominance of a peak at the very short wavelength limit of the emission spectrum. At 217 µW pump power, the prominent mode has blue-shifted to ~ 421 nm, and this microdisk sample ultimately exhibits lasing at that wavelength.

While the very broad gain spectrum reflects the optical emission of the entire range of fQWs and QDs that comprise the active layer, the dominant modes highlight the relative strengths of contribution of fQW and QDs as the incident pump power is changed. For the fQW+QD A and B samples, lasing ultimately occurs at wavelengths characteristic of the QDs, at wavelengths *much shorter* than the center of the room temperature gain spectrum. This distinctive behavior of these samples is observed over a significant sampling of microdisks as shown in Figure 4. A histogram of the wavelength of the lasing mode for each measured device is overlaid with the average background emission of the four samples. For fQW+QD A and B, the distribution of QD emission is also denoted. This distribution was determined by taking the average wavelength and standard deviation of QD exciton peaks, measured at low temperature. Finally, color bars represent the average lasing threshold of the microdisks within one histogram bin. While the lasing wavelengths for the QW and fQW devices are fairly equally distributed about the center-emission wavelengths for these materials, the lasing wavelengths of the fQW+QD materials are substantially blue-shifted (~ 13 nm for sample A) from the broadband gain emission and instead are centered on the QD emission spectrum.

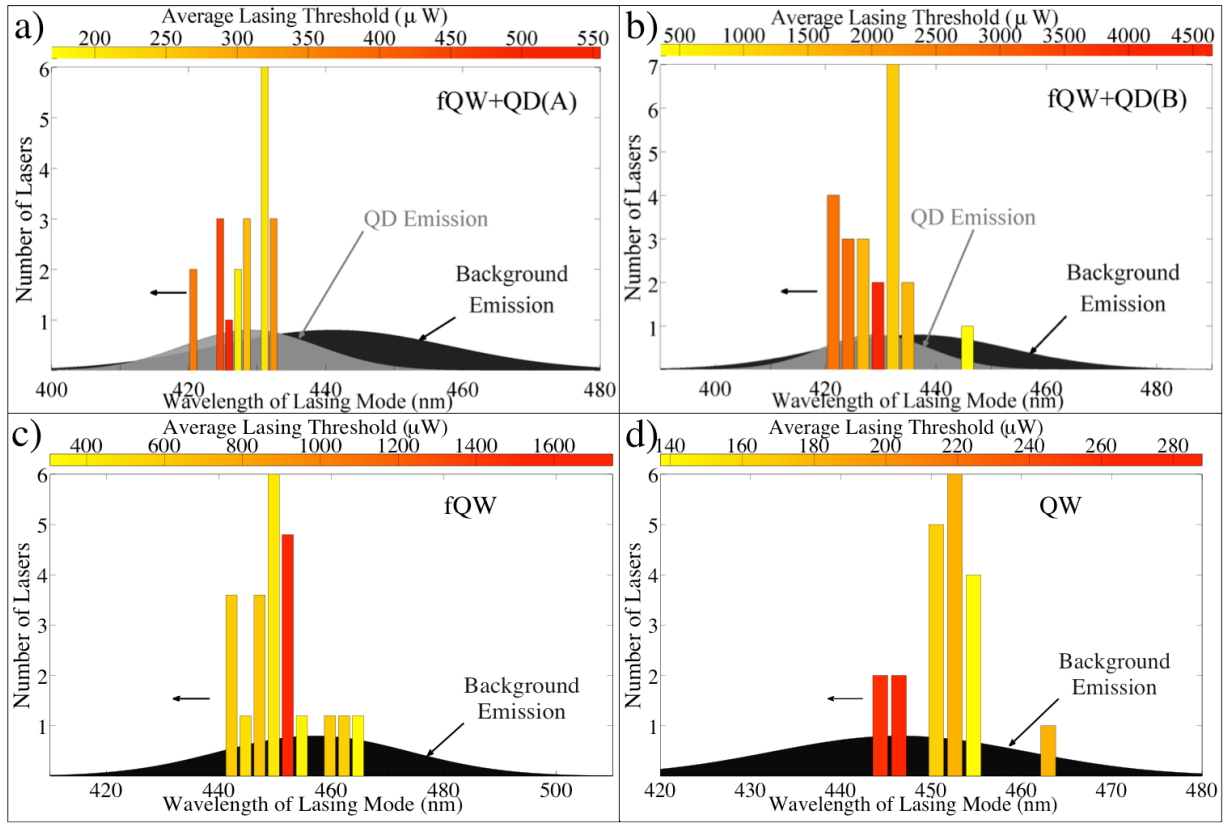

*Fig 4. Histogram of the wavelength of lasing mode for each laser measured from each of the four samples a) fQW+QD(A), b) fQW+QD(B), c) fQW, and d) QW. The average lasing threshold of the devices in each histogram bar is denoted by the color map, with yellow and red corresponding to low and high thresholds, respectively. The normalized background emission spectrum is denoted in black and the QD emission spectrum is denoted in gray for fQW+QD(A) and fQW+QD(B)*

For clarity, for each active layer composition, the averaged data for the center of the gain spectrum, the lasing wavelength, and the lasing threshold are also tabulated in Table I.

Lasing Statistics Summary

| Sample | Average wavelength of background emission (nm) | Average wavelength of lasing mode (nm) | Average Lasing Threshold (µW) | Threshold Range (µW) |
|---|---|---|---|---|
| QW | 446± 2.3 | 448± 6.7 | 184 | 88-375 |
| fQW | 457± 3.5 | 451± 6.7 | 753 | 83-3600 |
| fQW+QD A | 441± 4.8 | 428± 3.8 | 303 | 118-815 |
| fQW+QD B | 437± 4.4 | 429± 6.3 | 2029 | 350-6175 |

*Table I: Center of background emission spectrum, average wavelength of lasing mode, and lasing threshold determined by averaging over the 20 microdisk lasers measured for each of the 4 active layer materials.*

**Discussion**

The information on lasing thresholds and wavelengths for the various active areas, as well as the evolution of the spectra at differing pump powers provide significant insights into the important constraints on lasing in these microdisk devices. The evolution of the dominant mode from longer to shorter wavelengths as the pump power is increased reflects the interplay between and relative importance of the carrier capture cross-section versus the efficiency and rate of radiative emission of those fQW regions of varying size. The larger area fQW regions are more effective at carrier capture, with subsequent emission at the longer wavelengths characteristic of their energy levels. Efficient carrier capture may be the dominant advantage at the lowest pump powers. However because of better spatial localization of the excitons, the radiative efficiency and ultimately, the radiative rate will be higher for the smaller area fQWs (22). As the pump power increases, carrier capture may be less influential than the higher radiative efficiency emission rate of the smaller area fQWs and the emission spectrum shifts to the shorter wavelengths characteristic of the energy levels of those fQW regions. The behavior continues for the fQW+QD sample. At higher pump powers the QD emission into the mode at ~ 421 nm becomes observable, and ultimately lasing at that wavelength is observed. We note the relatively short wavelength exciton emission of the QDs in Figure 3, as well as the blue-shift of the gain spectrum as the pump power changed from 1.4 µW to 43 µW. Thus, QD lasing at 421 nm is consistent with the low temperature PL spectrum of excitonic transitions from the QDs. In addition, previous work suggests that the threshold carrier density for QDs is lower than that of QWs (23, 24). Thus at sufficiently high pump powers, the stronger spatial localization and higher radiative emission efficiency mediate the much lower relative carrier capture efficiency, and result in lasing at the short wavelengths characteristic of the QDs. Such a mechanism may also explain why fQW+QD (B) lases closer to the center of the background gain spectrum: its QD density is ~10x less than that of fQW+QD (A), and thus certain devices may not have the critical number of QDs to facilitate lasing. In this case the device behaves similarly to a fQW device.
In conclusion we have fabricated microdisk cavities containing QW, fQW, and fQW+QD layers as the gain medium. The modal signatures of the respective spectra show interplay between the large carrier capture cross section of larger area-fQWs at the lowest pump powers, yielding precedence to the higher carrier localization of smaller-area fQWs and QDs, at the higher pump powers. In fQW+QD devices we observe lasing at wavelengths centered on the QD gain spectrum, which is blue shifted from the peak of the broadband emission. We believe that this distinctive emission wavelength is a signature of QD lasing in these samples. These observations offer a way to distinguish lasing from QDs from that of the MDE-associated fQWs and provide insights into the optimal nitride gain medium in order to achieve lower laser thresholds and improved device performance.

**Materials and methods section**
Figure 5 shows AFM images of one uncapped active layer for each of the four samples. It is apparent that some portions of the QD+fQW samples contain fragmented QW material *or* QDs alone, while other portions of the material may comprise a QD overlying fQW material. Regions of the InGaN fQW, QD, and bare GaN are denoted in figure 1 (b).

AFM scans of the InGaN layers were performed on uncapped versions of the samples. The QW sample (Fig 5a) shows a smooth top surface of the InGaN QW (RMS roughness of 0.16 nm), whereas the other sample scans (Fig 5 b,c,d) show areas of both underlying GaN and the fQW surfaces. The depth of the fQW has been measured as 2.5 nm for all fQW samples.

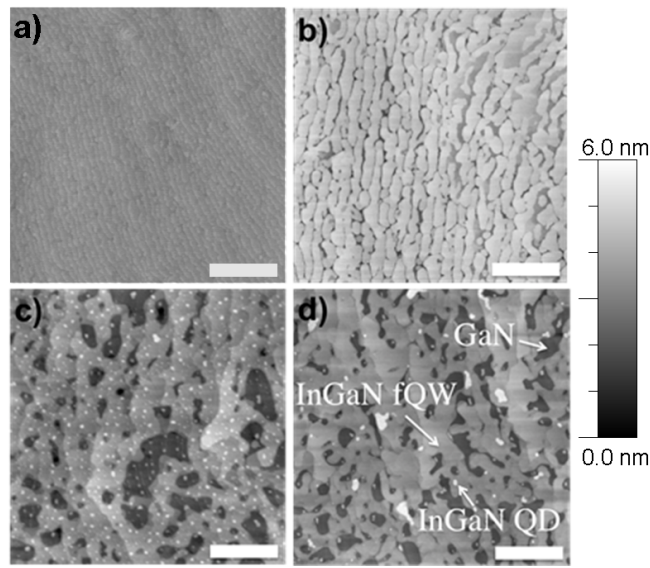

*Fig 5: AFM images of uncapped active layers a) QW sample showing atomic terraces of the top surface of the InGaN QW, b) fQW, c) QD+fQW A and d) QD+fQW B. Scale bar is 500nm in width, with a vertical scale of 6 nm. The white bright dots in c) and d) are indium droplets which form InGaN QDs during growth of the GaN capping layer. The dark patches in b), c) and d) are bare GaN.*

The details of the growth methods for the active regions are as follows:

*InGaN QWs* – To minimize the indium loss from the QW during barrier growth, GaN barriers were grown at the same temperature as the InGaN QWs. The nominal thickness for the QWs is 2.5 nm. This method results in continuous QWs with no visible well width fluctuation measured using Transmission Electron Microscopy (TEM) (25).

*InGaN fQWs* – 2.5 nm thick InGaN epilayers were grown and annealed at the growth temperature for 240 seconds in an $NH_3/N_2$ atmosphere prior to the growth of a GaN capping layer. The use of $NH_3/N_2$ gas mixture was to ensure that no metallic droplets are formed during the annealing process. The annealed epilayer exhibits a network of interlinking InGaN strips aligned along the [11-20] direction as shown in Figure 1 (a) (26).

*$In_xGa_{1-x}N$ QDs + $In_xGa_{1-x}N$ fQWs* – Modified Droplet Epitaxy (MDE) was used for the growth of InGaN QDs in our samples. The full MDE technique is discussed elsewhere (27), however a brief description follows: a post-growth $N_2$ anneal is performed on a 2.5 nm InGaN QW layer causing the layer to be broken into regions of 'pits' and regions of 'fragmented QW', with metallic indium/gallium droplets being created across both of these regions. During the growth of the GaN capping layer, these droplets re-react with ammonia, forming InGaN QDs.

**Acknowledgements**
This work was supported in part by the Engineering and Physical Sciences Research Council (Award No. EP/H047816/1) and by the NSF Materials World Network (Award No. 1008480). This work was enabled by facilities available at the Center for Nanoscale Systems, a member of the National Nanotechnology Infrastructure Network (NNIN) which is supported by the National Science Foundation under NSF Award No. ECS-0335765. Dr. Aharonovich is the recipient of an Australian Research Council Discovery Early Career Research Award (Project No. DE130100592).